\documentclass[aps,prd,twocolumn,nofootinbib,floatfix]{revtex4} % {revtex4-1}

\usepackage{graphics}
\usepackage{graphicx}

\usepackage{latexsym}
\usepackage[cp866]{inputenc}
\usepackage[english]{babel}

\input epsf

\begin{document}

\title{\boldmath
Decay $X(3872)\to\pi^0\pi^+\pi^-$ and $S$-wave $D^0\bar D^0
\to\pi^+\pi^-$ scattering length}
\author{N.~N.~Achasov and G.~N.~Shestakov}
\affiliation{Laboratory of Theoretical Physics, S.~L.~Sobolev
Institute for Mathematics, 630090 Novosibirsk, Russia}

\begin{abstract}
The isospin-breaking decay $X(3872)\to(D^*\bar D+ \bar D^*D)
\to\pi^0D\bar D\to\pi^0\pi^+\pi^-$ is discussed. In its amplitude
there is a triangle logarithmic singularity, due to which the
dominant contribution to $BR(X(3872)\to\pi^0\pi^+\pi^-)$ comes from
the production of the $\pi^+\pi^-$ system in a narrow interval of
the invariant mass $m_{\pi^+\pi^-}$ near the value of $2m_{D^0}
\approx 3.73$ GeV. The analysis shows that $BR(X(3872)\to\pi^0
\pi^+\pi^-)$ can be expected at the level of $10^{-3}$--$10^{-4}$.
This estimate includes, in particular, the assumption that the
$S$-wave inelastic scattering length $|\alpha''_{D^0\bar D^0
\to\pi^+\pi^-}|\approx1/(2m_{D^{*+}})\approx0.25\ \mbox{GeV}^{-1}$.
\end{abstract}

\maketitle

\section{Introduction}

The state $X(3872)$ [or $\chi_{c1}(3872)$ \cite{PDG18}] was first
observed in 2003 by the Belle Collaboration in the process $B\to
K(X(3872)\to\pi^+\pi^-J/\psi)$ \cite{Cho03}. Then it was observed in
many other experiments in other processes and decay channels
\cite{PDG18}. The $X(3872)$ is a narrow resonance in
non-$(D^{*0}\bar D^0+\bar D^{*0} D^0)$ decay channels,
$\Gamma_X<1.2$ MeV \cite{Cho11}, and its mass coincides practically
with the $D^{*0}\bar D^0$ threshold \cite{PDG18}. It has the quantum
numbers $I^G(J^{PC})=0^+(1^{++})$ \cite{PDG18,Aub05,Aai15}. In
addition to decay into $\pi^+\pi^-J/\psi$ \cite{Cho03,Aai13,Abl14},
the $X(3872)$ also decays into $\omega J/\psi$ \cite{Abe05,Amo10,
Abl19,FN1}, $D^{*0}\bar D^0+c.c.$ \cite{Gok06,Aus10}, $\gamma
J/\psi$ \cite{Aub09,Bha11, Aai14}, $\gamma\psi(2S)$
\cite{Aub09,Bha11,Aai14}, and $\pi^0\chi_{c1}(1P)$ \cite{Ab19,FN2}.
The nature of $X(3872)$ remains the subject of much discussion; see,
for example, Refs. \cite{Aub09,Bha11,Aai14,FN2,
Ab19,Che16,Esp16,Leb17,Ali17,Ols18,Guo18,AR14,AR15,AR16,A16,A17,
Ach18,Mut,Sun,Fer}. Of course, new experiments will allow making a
more definite choice between different interpretations.

The search for $X(3872)$ in decay channels that do not contain
charmed particles or charmonium states [i.e., in channels other than
$D^{*0}\bar D^0+c.c.$, $D^0\bar D^0\pi^0$, $\pi^+\pi^-J/\psi$,
$\omega J/\psi$, $\gamma J/\psi$, $\gamma\psi (2S)$, $\pi^+\pi^-
\eta_c(1S)$, $\pi^+\pi^-\chi_{c1}(1P)$, and $\pi^0\chi_{c1}(1P)$] is
of great interest \cite{PDG18,Ach18,Mut,Sun,Fer,AR15,AR16,A16,
A17,AR14,Aai17,Bar18}. For example, the $c\bar c=\chi_{c1}(2P)$
scenario predicts a significant number of various two gluon decays
$X(3872)\to(\mbox{gluon}+\mbox{gluon})\to \mbox{light\ hadrons}$
\cite{Ach18,AR15,AR16,A16,A17}. The situation here is qualitatively
the same as for the decays
$\chi_{c1}(1P)\to(\mbox{gluon}+\mbox{gluon})\to \mbox{light\
hadrons}$. In this way, only one channel has been explored so far
\cite{PDG18}. Namely, the LHCb Collaboration undertook a search for
the decay $X(3872)\to p\bar p$, which resulted in the following
restriction \cite{Aai17}:
\begin{eqnarray}\label{Eq1a}
\frac{BR(B^+\to X(3872)K^+)\times(BR(X(3872)\to p\bar p)}{
BR(B^+\to J/\psi K^+)\times(BR(J/\psi\to p\bar p)} \nonumber \\
<0.25\times10^{-2}.\qquad\qquad\qquad\end{eqnarray} Hence, in view
of $ BR(B^+\to J/\psi K^+)\times(BR(J/\psi\to p\bar p)\approx2.14
\times10^{-6}$ \cite{PDG18} and $0.9\times10^{-4}<BR(B^+\to
X(3872)K^+)<2.7\times10^{-4}$ \cite{PDG18,Yua}, it follows that
\begin{eqnarray}\label{Eq2a}
BR(X(3872)\to p\bar p)<0.6 \times10^{-4}.\end{eqnarray} Taking into
account a sizable contribution of the $D^{*0}\bar D^0+\bar D^{*0}
D^0$ channel (and also the channels containing the charmonium
states) to the $X(3872)$ decay rate, one can conclude that the above
relation is in satisfactory agreement (at least not in
contradiction) with what is observed in the decays of the
$\chi_{c1}(1P)$ meson: $BR(\chi_{c1}(1P)\to p\bar p)=(7.60\pm0.34)
\times10^{-5}$ \cite{PDG18}. Note that the $\chi_{c1}(1P)$ has only
one decay into $\gamma J/\psi$ containing $c\bar c$ quarks in the
final state. It is also proposed to investigate the $X(3872)$
coupling to the $p\bar p$ channel in the reaction $p\bar p\to
X(3872)\to\pi^+\pi^-J/\psi$ with the PANDA detector \cite{Bar18}.

We propose to obtain an experimental limit on the probability of the
decay $X(3872)\to \pi^0\pi^+\pi^-$ and, if lucky, to register this
decay. According to our estimate, the branching ratio of the decay
$X(3872)\to\pi^0 \pi^+\pi^-$ can be expected at the level of
$10^{-3}$--$ 10^{-4}$ due to the transition mechanism
$X(3872)\to(D^*\bar D+\bar D^*D)\to\pi^0D\bar D\to\pi^0\pi^+\pi^-$.
In this case, the main contribution to
$BR(X(3872)\to\pi^0\pi^+\pi^-)$ comes from the production of
$\pi^+\pi^-$ pairs in a narrow interval of the invariant mass
$m_{\pi^+\pi^-}$ near the value of $2m_{D^0} \approx 3.73$ GeV.

As for the nature of X(3872), our calculations implicitly imply for
this state the conventional $c\bar c$ nature, i.e., that it is a
compact charmonium state similar to the states $\chi_{c1}(1P)$,
$\psi(2S)$, $\psi(3770)$, and so on, and to describe its decays one
can use the effective phenomenological Lagrangian approach
\cite{AR14,AR15,AR16,A16,A17,Ach18}.

\section{Estimate of \boldmath $BR(X(3872)\to\pi^0\pi^+\pi^-)$}

The decay $X(3872)\to(D^{*0}\bar D^0+\bar D^{*0}D^0)\to\pi^0D^0\bar
D^0$ (see Fig. 1) is one of the main decay channels of the $X(3872)$
resonance \cite{PDG18}. Because of the final state interaction among
$D^0$ and $\bar D^0$ mesons, i.e., due to the $S$-wave transition
$D^0\bar D^0\to\pi^+\pi^-$, the isospin breaking decay
$X(3872)\to(D^{*0}\bar D^0+\bar D^{*0}D^0) \to\pi^0D^0\bar
D^0\to\pi^0\pi^+\pi^-$ is induced (see Fig. 2).

The amplitudes of such triangle diagrams, as in Fig. 2, may contain
logarithmic singularities that can produce some enhancement in the
mass spectra. The conditions for the appearance of such
singularities in the physical region of the reaction were repeatedly
deduced in various forms and discussed in the literature; see, for
example, Refs. \cite{KSW58,Lan59,FN64,Val64,Ait64,CN65,Mik15,Liu16,
Bay16} and also the very recent work \cite{Guo19}. For the
considered mechanism of the $X\to\pi^0\pi^+\pi^-$ decay, these
conditions are reduced to the following relations.

%--------------------------------------------------------------------------------
\begin{figure} %[!ht]
\begin{center}\includegraphics[width=7cm]{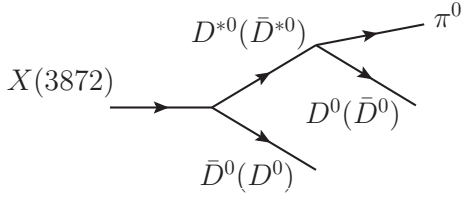}
\caption{\label{Fig-X-3} The diagram of the decay $X(3872)\to D^0
\bar D^0\pi^0$. The four-momenta of $X(3872)$, $D^0$, $\bar D^0$,
and $\pi^0$ are, respectively, $p_1$, $p_D$, $p_{\bar D}$, and
$p_\pi$; the four-momenta of the intermediate $D^{*0}$ and $\bar
D^{*0}$ are $k_1$ and $k_2$, respectively.}\end{center}\end{figure}
%--------------------------------------------------------------------------------

%--------------------------------------------------------------------------------
\begin{figure} %[!ht]
\begin{center}\includegraphics[width=7cm]{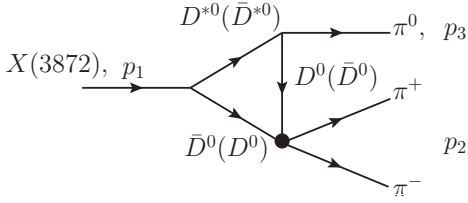}
\caption{\label{Fig-X-1} The diagram of the decay $X(3872)\to
(D^{*0}\bar D^0+\bar D^{*0}D^0)\to\pi^0D^0\bar D^0\to\pi^0\pi^+
\pi^-$. In the $X(3872)$ mass region, all intermediate particles in
the triangle loop can be near or directly on the mass shell. As a
consequence, a logarithmic singularity in the imaginary part of the
amplitude emerges in the hypothetical case of the stable $D^{*0}$
meson when the conditions (\ref{Eq3a}) and (\ref{Eq4a}) are
fulfilled. The four-momenta of corresponding particles are denoted
as $p_1$, $p_2$, and $p_3$; $p_1^2=s_1$ is the squared invariant
mass of the $X(3872)$ resonance or of the final $\pi^0\pi^+\pi^-$
system; $p_2^2=s_2=m^2_{\pi^+\pi^-}$ is the squared invariant mass
of the final $\pi^+\pi^-$ system; and $p_3^2=m^2_{\pi^0}$.
}\end{center}\end{figure}
%--------------------------------------------------------------------------------

If the virtual invariant mass squared of the $X(3872)$ resonance
$s_1$ falls in the range
\begin{eqnarray}\label{Eq3a}
2(m^2_{D^{*0}}+m^2_{D^0})-m^2_{\pi^0}=(3.87193\ \mbox{GeV})^2
>s_1\nonumber\\ >(m_{D^{*0}}+m_{D^0})^2=(3.87168\ \mbox{GeV})^2
,\qquad\end{eqnarray} then, in the range of the invariant mass
squared of the $\pi^+\pi^-$ system $s_2=m^2_{ \pi^+\pi^-}$
\begin{eqnarray}\label{Eq4a}
\frac{m_{D^0}}{m_{D^{*0}}}(m^2_{D^{*0}}+m^2_{D^0}-m^2_{\pi^0})
+2m^2_{D^0}=(3.7299\ \mbox{GeV})^2\nonumber\\
>s_2>4m^2_{D^0}=(3.72966\ \mbox{GeV})^2,\qquad\qquad
\end{eqnarray}
the imaginary part of the amplitude of the diagram in Fig. 2
contains the triangle logarithmic singularity
\cite{KSW58,Lan59,FN64,Val64,Ait64,CN65,Mik15,Liu16,Bay16,Guo19}.
Below, we see that this singularity leads to the resonancelike
enhancement in the $\pi^+\pi^-$ mass spectrum at
$\sqrt{s_2}=m_{\pi^+\pi^-} \approx 2m_{D^0}\approx3.73$ GeV, i.e.,
near the $D^0\bar D^0$ threshold.

The decay $X(3872)\to\pi^0\pi^+\pi^-$ can also be produced via the
charged intermediate states, $X(3872)\to(D^{*+}D^-+D^{*-}D^+)
\to\pi^0D^+D^-\to\pi^0\pi^+\pi^-$ (see Fig. 3).
%--------------------------------------------------------------------------------
\begin{figure} %[!ht]
\begin{center}\includegraphics[width=7cm]{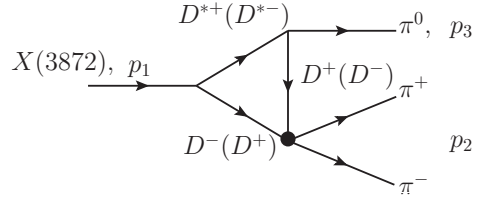}
\caption{\label{Fig-X-4} The diagram of the decay $X(3872)\to\pi^0
\pi^+\pi^-$ corresponding to the charged intermediate state
contributions, $X(3872)\to(D^{*+}D^-+D^{*-}D^+)\to\pi^0D^+D^-
\to\pi^0\pi^+\pi^-$.} \end{center}\end{figure}
%--------------------------------------------------------------------------------
From the isotopic symmetry for the coupling constants (C invariance
of the amplitudes is implied), it follows that the contributions of
the diagrams in Figs. 2 and 3 exactly compensate each other and the
isospin breaking decay $X(3872)\to \pi^0\pi^+\pi^-$ is absent, if
$m_{D^{*+}}=m_{D^{*0}}$ and $m_{D^+}=m_{D^0}$. However, the
$D^{*0}\bar D^0$ and $D^{*+}D^-$ thresholds in the variable
$\sqrt{s_1}$ differ by 8.23 MeV ($m_{D^{*0}}+ m_{\bar D^0}=3.87168$
GeV, $m_{D^{*+}}+ m_{D^-}=3.87991$ GeV) and the $D^{0}\bar D^0$ and
$D^{+}D^-$ thresholds in the variable $\sqrt{s_2}$ differ by 9.644
MeV ($2m_{D^0}=3.72966$ GeV, $2m_{D^\pm}=3.73930$ GeV). Therefore,
in the region of the variables $\sqrt{s_1}$ and $\sqrt{s_2}$ that is
significant for the decay $X(3872)\to\pi^0\pi^+ \pi^-$ (i.e., for
$\sqrt{s_1}\approx m_X\approx m_{D^{*0}}+m_{\bar D^0}$, where $m_X$
is the nominal mass of the $X(3872)$ equal to 3.87169 GeV
\cite{PDG18}, and $\sqrt{s_2}\approx2m_{\bar D^0}\approx3.73$ GeV),
the contributions from the neutral (see Fig. 2) and charged (see
Fig. 3) intermediate states weakly compensate each other and the
contribution of the diagram in Fig. 2 dominates.
%--------------------------------------------------------------------------------

We write the differential probability for the decay of the virtual
state $X(3872)$ to $\pi^0\pi^+\pi^-$ in the form
\begin{eqnarray}\label{Eq5a}
\frac{d^2BR(X\to\pi^0\pi^+\pi^-;s_1,s_2)}{d\sqrt{s_1}d\sqrt{s_2}}
\qquad\quad\nonumber \\ =\frac{2
\sqrt{s_1}}{\pi}\frac{\sqrt{s_1}}{|D_{X}(s_1)|^2}\frac{d\Gamma(X\to
\pi^0\pi^+\pi^-;s_1,s_2)}{d\sqrt{s_2}},\end{eqnarray} where $D_{X}
(s_1)$ is the inverse propagator of the $X(3872)$ resonance
\cite{AR14,AR16,A16} that takes into account the couplings of
$X(3872)$ with the $D^*\bar D+\bar D^* D$ decay channels as well as
with all non-($D^*\bar D+\bar D^*D $) decay channels; and
$d\Gamma(X\to \pi^0\pi^+\pi^-;s_1,s_2)/d \sqrt{s_2}$ is the
$X\to\pi^0 \pi^+\pi^-$ differential decay width in the variable
$\sqrt{s_2}=m_{\pi^+\pi^-}$ caused by the sum of the diagrams in
Figs. 2 and 3.

%--------------------------------------------------------------------------------
The $X(3872)$ resonance propagator constructed in Refs. \cite{AR14,
AR16,A16} has good analytical and unitary properties. The inverse
propagator $D_X(s_1)$ has the form \cite{AR14,AR16,A16}
\begin{eqnarray}\label{Eq24a}
D_X(s_1)=m^2_X-s_1\qquad\qquad\nonumber\\ +\sum_{ab}
[\mbox{Re}\Pi^{ab}_X(m^2_X)-\Pi^{ab}_X(s_1)]-im_X\Gamma_{non},
\end{eqnarray}
where $\Gamma_{non}=\Sigma_i\Gamma_i$ is the total width of the
$X(3872)$ decay to all non-$(D^*\bar D+\bar D^*D)$ channels which in
the narrow region of the $X(3872)$ peak ($\Gamma_X<1.2$ MeV
\cite{PDG18,Cho11}) is approximated by a constant; $ab=D^{^*0}\bar
D^0$, $\bar D^{^*0}D^0$, $D^{^*+}D^-$, $D^{^*-}D^+$. At
$s_1>(m_a+m_b)^2$
\begin{eqnarray}\label{Eq25a}\Pi^{ab}_{X}(s_1)=\frac{g^2_{A}}
{16\pi}\left[\frac{m_{ab}^{(+)}m_{ab}^{(-)}}{\pi
s_1}\ln\frac{m_b}{m_a}+\rho_{ab}(s_1)\right.\qquad\nonumber\\
\left.\times\left(i-\frac{1}{\pi}\,\ln\frac{\sqrt{s_1-m_{ab}^{(-)
\,2}}+\sqrt{s_1-m_{ab}^{(+)\,2}}}{\sqrt{s_1-m_{ab}^{(-)\,2}}-\sqrt{s_1
-m_{ab}^{(+)\,2}}}\right)\right],\ \ \end{eqnarray} where
$\rho_{ab}(s_1)=\sqrt{s_1-m_{ab}^{(+)\,2}}\,\sqrt{s_1-m_{ab}^{(-)
\,2}}/s$, $m_{ab}^{(\pm)}=m_a\pm m_b$, $m_a>m_b$,
\begin{eqnarray}\label{Eq26a}
\mbox{Im}\,\Pi^{ab}_X(s_1)=\sqrt{s_1}\Gamma_{X\to ab}
(s_1)=\frac{g^2_A}{16\pi}\rho_{ab}(s_1),\end{eqnarray} and $g_A$ is
the coupling constant of $X$ with the $D^{*0}\bar D^0$ channel.
%--------------------------------------------------------------------------------
At $m_{ab}^{(-)\,2}<s_1<m_{ab}^{(+)\,2}$
\begin{eqnarray}\label{Eq27a}\Pi^{ab}_{X}(s_1)=\frac{g^2_{A}}
{16\pi} \left[\frac{m_{ab}^{(+)}m_{ab}^{(-)}}{\pi
s_1}\ln\frac{m_b}{m_a}\right.\qquad\nonumber\\ \left.
-\rho_{ab}(s_1)\left(1-\frac{2}{\pi}\arctan\frac{\sqrt{
m_{ab}^{(+)\,2}-s_1}}{\sqrt{s_1-m_{ab}^{(-)\,2}}}\right)\right],
\end{eqnarray}
where  $\rho_{ab}(s_1)$\,=\,$\sqrt{m_{ab}^{(+)\,2}-s_1}
\,\sqrt{s_1-m_{ab}^{(-)\,2}}\,/s_1$. If $s_1\leq m_{ab}^{(-)\,2}$,
then $\rho_{ab}(s_1)$\,=\,$\sqrt{m_{ab}^{(+)\,2}-s_1}\,\sqrt{
m_{ab}^{(-)\,2}-s_1}\,/s_1$, and
\begin{eqnarray}\label{Eq28a}\Pi^{ab}_{X}(s_1)=\frac{g^2_{A}}
{16\pi}\left[\frac{m_{ab}^{(+)}m_{ab}^{(-)}}{\pi
s_1}\ln\frac{m_b}{m_a}\right.\qquad\quad\nonumber\\ \left.
+\rho_{ab}(s_1)\frac{1}{\pi}\,\ln\frac{
\sqrt{m_{ab}^{(+)\,2}-s_1}+\sqrt{m_{ab}^{(-)\,2}-s_1}}
{\sqrt{m_{ab}^{(+)\,2}-s_1}-\sqrt{m_{ab}^{(-)\,2}-s_1}}\right].
\end{eqnarray}
The sum of the probabilities of the $X(3872)$ decay to all modes
satisfies the unitarity \cite{AR14,AR16,A16}
\begin{eqnarray}\label{Eq29a}
BR(X\to(D^{*0}\bar D^0+c.c.))\qquad\qquad\ \nonumber \\
+ BR(X\to(D^{*+}D^-+c.c))+\Sigma_i BR(X\to i)=1.
\end{eqnarray}
The coupling of the $X(3872)$ with the $D^{*0}\bar D^0$ system was
introduced in Refs. \cite{AR14,AR15,AR16,A16} by means of the
Lagrangian
\begin{eqnarray}\label{Eq30a}
L_{XD^{*0}\bar D^0}(x)=g_A X^\mu(D^{*0}_\mu\bar D^0+\bar D^{*0}_\mu
D^0)\end{eqnarray} and the range of possible values of the coupling
constant $g^2_A/(16\pi)$ was determined from the analysis of the
experimental data \cite{Cho11,Abe05,Amo10,Aus10,Aai13,Bha11}.

To describe the amplitudes of the $D^*\to D\pi^0$ decays, we use the
expression
\begin{eqnarray}\label{Eq12a}
V_{D^*D\pi^0}=g_{D^*D\pi^0}(\epsilon_{D^*},p_{\pi^0}-p_D)\,,
\end{eqnarray} where $\epsilon_{D^*}$ is the polarization
four-vector of the $D^*$ meson, $p_{\pi^0}$ and $p_D$ are the
four-momenta of $\pi^0$ and $D$, respectively; $\,g_{D^{*+}D^+
\pi^0}=-g_{D^{*0}D^0\pi^0}$.

The effective vertex of the $X(3872)\to(D^*\bar D+\bar
D^*D)\to\pi^0D\bar D\to\pi^0\pi^+\pi^-$ transition corresponding to
the sum of the diagrams in Figs. 2 and 3, in which the $\pi^+\pi^-$
system is produced in the $S$ wave, can be written as
\begin{eqnarray}\label{Eq5-3-6}
V_{X\pi^0\pi^+\pi^-}=G_{X\pi^0\pi^+\pi^-}(s_1,s_2)(\epsilon_X,p_3-p_2)\nonumber\\
=2\frac{\bar{g}}{16\pi}[F_0(s_1,s_2) -F_+(s_1,s_2)]\,,\qquad
\end{eqnarray} where the invariant amplitude
$G_{X\pi^0\pi^+\pi^-}(s_1,s_2)$ is used below [see, Eq.
(\ref{Eq9a})] to compactly write the expression for the energy
dependent differential width of the $X\to\pi^0\pi^+\pi^-$ decay;
$\epsilon_X$ is the polarization four-vector of the $X(3872)$, the
amplitudes $F_0(s_1,s_2)$ and $F_+(s_1,s_2)$ describe the
contributions from the neutral and charged intermediate $D^*\bar D$
states, respectively, and
\begin{eqnarray} \label{Eq16a} \bar{g}_=g_A\,g_{D^{*0}D^0\pi^0}\,g_{D^0\bar
D^0\pi^+\pi^-}.\end{eqnarray} We assume the $S$-wave amplitudes of
the processes $D^0\bar D^0\to\pi^+\pi^-$ and $D^+\bar
D^-\to\pi^+\pi^-$ (entering in the amplitudes of the diagrams in
Figs. 2 and 3) to be equal and approximate them in the region of the
$D\bar D$ thresholds by an $s_2$-independent constant $g_{D^0\bar
D^0\pi^+\pi^-}$.

Taking into account Eqs. (\ref{Eq30a})--(\ref{Eq5-3-6}), the
amplitude $F_0(s_1,s_2)$ can be written in the form
\begin{eqnarray} \label{Eq17b} F_0(s_1,s_2)=\frac{i}{\pi^3}\,
\epsilon_{X\mu}\int \frac{\left(-g_{\mu\nu}+\frac{k_\mu
k_\nu}{m^2_{D^{*0}}}\right)(2p_{3\nu}-k_\nu)}{(k^2-
m^2_{D^{*0}}+i\varepsilon)}\quad \nonumber\\ \times\,
\frac{d^4k}{((p_1-k)^2-m^2_{D^0}+i\varepsilon)((k-p_3)^2-m^2_{\bar
D^0}+i\varepsilon)}.\ \ \end{eqnarray} The four-vector under the
integral sign we transform as follows
\begin{eqnarray} \label{Eq17c}
\left(-g_{\mu\nu}-\frac{k_\mu k_\nu}{m^2_{D^*}}\right)(2p_{3\nu}
-k_\nu)=-2p_{3\mu}+k_\mu\left(m^2_{D^{*0}}\right.\nonumber\\
\left.-m^2_{D^0}+m^2_{\pi^0}\right)/m^2_{D^{*0}}-k_\mu((k-p_3)^2-m^2_{\bar
D^0})/m^2_{D^{*0}}.\ \end{eqnarray} This shows that after reducing
the numerator and denominator in Eq. (\ref{Eq17b}) by the factor
$((k-p_3)^2-m^2_{\bar D^0})$, the divergent part of the integral is
proportional to $p_{1\mu}$ [i.e., the four-moment of the $X(3872)$
resonance] and does not contribute to $F_0(s_1,s_2)$ because
$(\epsilon_X,p_1)=0$. For the numerical calculation of the
amplitudes $F_0(s_1,s_2)$ in Eq. (16), we use the method developed
in Refs. \cite{tHV79,PV79}. Note that the part of the contribution
from the second term in (17), $k_\mu(m^2_{D^{*0}}-
m^2_{D^0}+m^2_{\pi^0})/ m^2_{D^{*0}}$, which after integration turns
out to be proportional to $p_{3\mu}$, gives a negligible
contribution to $F_0(s_1,s_2)$ in the $\sqrt{s_1}$ and $\sqrt{s_2}$
region under consideration. Thus we put
%--------------------------------------------------------------------------------------------
\begin{eqnarray} \label{Eq17d} F_0(s_1,s_2)=-2(\epsilon_X,p_{3})
\frac{i}{\pi^3}\,\int \frac{d^4k}{(k^2- m^2_{D^{*0}}+i\varepsilon)}
\quad\ \, \nonumber\\ \times\frac{1} {((p_1-k)^2-m^2_{D^0}+i
\varepsilon)((k-p_3)^2-m^2_{\bar D^0}+i\varepsilon)}.\ \
\end{eqnarray} The amplitude $F_+(s_1,s_2)$ is obtained from Eq.
(\ref{Eq17d}) by replacing the masses of neutral $D^*$ and $D$
mesons by the masses of their charged partners.
%--------------------------------------------------------------------------------------------

Using Eq. (\ref{Eq5-3-6}) we express the differential width
$d\Gamma(X\to\pi^0\pi^+\pi^-;s_1,s_2)/d\sqrt{s_2}$ in terms of the
invariant amplitude $G_{X\pi^0\pi^+\pi^-}(s_1,s_2)$.
\begin{eqnarray}\label{Eq9a}
\frac{d\Gamma(X\to\pi^0\pi^+\pi^-;s_1,s_2)}{d\sqrt{s_2}}\qquad\qquad\ \ \nonumber\\
=\frac{2}{3}\,\frac{|G_{X\pi^0\pi^+\pi^-}(s_1,s_2)|^2}{4\pi}\,\frac{p^3(s_1,s_2)}{
s_1}\,\frac{\rho(s_2)}{16\pi}\,\frac{2\sqrt{s_2}}{\pi}\,,\end{eqnarray}
where
\begin{eqnarray}\label{Eq10a}
p(s_1,s_2)=\frac{\sqrt{s_1^2-2s_1(s_2+m^2_{\pi^0})+(s_2-m^2_{\pi^0})^2}}{2\sqrt{s_1}},
\\ \label{Eq10b} \rho(s_2)=\sqrt{1-4m^2_{\pi^+}/s_2}.
\qquad\qquad\quad\end{eqnarray} The width of the decay
$X\to\pi^0\pi^+\pi^-$ as a function of $s_1$ has the form
\begin{eqnarray}\label{Eq6a}
\Gamma(X\to\pi^0\pi^+\pi^-;s_1)\qquad\qquad\nonumber\\
=\int^{\sqrt{s_1}-m_\pi^0}_{2m_{\pi^+}}
\frac{d\Gamma(X\to\pi^0\pi^+\pi^-;s_1,s_2)}{d\sqrt{s_2}}d\sqrt{s_2}\,,
\end{eqnarray} and the probability of this decay is given by the
expression
\begin{eqnarray}\label{Eq7a}
BR(X\to\pi^0\pi^+\pi^-)\qquad\qquad\ \nonumber
\\=\int^\infty_{3m_\pi}\frac{2\sqrt{s_1}}{\pi}\,\frac{\sqrt{s_1}\Gamma(X\to
\pi^0\pi^+\pi^-;s_1)}{|D_{X}(s_1)|^2}d\sqrt{s_1}\,.\end{eqnarray}
Equations (\ref{Eq6a}) and (\ref{Eq7a}) indicate the kinematically
allowable limits of integration. In fact, the main contributions in
Eqs. (\ref{Eq6a}) and (\ref{Eq7a}) are concentrated in much smaller
intervals.

%--------------------------------------------------------------------------------
We now estimate the coupling constants $g_{D^{*0}D^0\pi^0}$ and
$g_{D^0\bar D^0\pi^+\pi^-}$.

For the total decay width of the $D^{*0}$ meson, only its upper
limit is known so far: $\Gamma_{D^{*0}}<2.1$ MeV \cite{PDG18}. On
the other hand, the total decay width of the $D^{*+}$ meson and the
branching ratio of the $D^{*+}\to(D\pi)^+$ decay are well known
\cite{PDG18}: $\Gamma_{D^{*+}} \approx 83.6$ keV, $BR(D^{*+}\to
(D\pi)^+)\approx98.4\%$. Assuming the isotopic symmetry for the
coupling constants $g_{D^*D\pi}$, we have
\begin{eqnarray}\label{Eq22a}
\frac{m^2_{D^{*0}}\Gamma_{D^{*0}\to D^0\pi^0}}{p^3_{
D^0\pi^0}}=\frac{m^2_{D^{*+}}\Gamma_{D^{*+}\to(D\pi)^+}}
{2p^3_{D^0\pi^+}+p^3_{D^+\pi^0}}\,,
\end{eqnarray} where $p_{D\pi}$ denotes the momentum of the final $D$ or
$\pi$ meson in the $D^{*}$ rest frame. From here we find the decay
width $\Gamma_{D^{*0}\to D^0\pi^0}\approx36\,\mbox{keV}$ and the
coupling constant $g^2_{D^{*0}D^0\pi^0}/(4\pi)=3m^2_{D^{*0}}
\Gamma_{D^{*0}\to D^0\pi^0}/(2p^3_{D^0\pi^0})\approx2.8$. Using also
the value of $BR(D^{*0}\to D^0\pi^0)\approx64.7\%$ \cite{PDG18}, we
get an estimate for the total decay width of the $ D^{*0}$ meson:
$\Gamma_{D^{*0}}\approx55.6$ keV. Here we note in passing the
following. As the examples \cite{AK1,AK2,AK3,AKS15,AS17,AS18} show,
the instability of the vector mesons in the intermediate states
(i.e., the finiteness of their total widths) is important to take
into account when estimating the contributions of logarithmic
triangle singularities. In this case, $\Gamma_{D^{*0}}$ is small.
Nevertheless, its accounting in the $D^{*0}$ propagator (by
replacing $m^2_{D^{*0}}\to m^2_{D^{*0}}-im_{ D^{*0}}\Gamma_{
D^{*0}}$) noticeably smoothes the logarithmic singularity in the
amplitude of the diagram in Fig. 2 and the computed width $\Gamma
(X(3872)\to\pi^0\pi^+\pi^-;m_X)$ is reduced by approximately 30\% as
compared to that for $\Gamma_{D^{*0 }}=0$. In a similar way, we take
into account the width $\Gamma_{D^{*\pm}}$ in the $D^{*\pm}$
propagator.

The constant $g_{D^0\bar D^0\pi^+\pi^-}$ is associated with the
annihilation cross section $\sigma_{D^0\bar D^0\to\pi^+\pi^-}$ at
the $D^0\bar D^0$ threshold and with the corresponding inelastic
scattering length $\alpha''_{D^0\bar D^0\to\pi^+\pi^-}$ by the
relations:
\begin{eqnarray}\label{Eq23a}
\frac{k\,\sigma_{D^0\bar D^0\to\pi^+\pi^-}}{4\pi}=|\alpha''_{D^0\bar
D^0\to\pi^+\pi^-}|=q\left|\frac{g_{D^0\bar D^0\pi^+\pi^-}}{
8\pi\sqrt{s_2}}\right|^2,\ \end{eqnarray} where $k$ and $q$ are
momenta of the $D^0$ and $\pi^+$ mesons, respectively, in the
center-of-mass frame of the reaction $D^0\bar D^0\to\pi^+\pi^-$. In
the $D^0\bar D^0$ threshold domain of interest to us,
$q/s_2\approx1/(4m_{D^0})$. At present, the values in Eq.
(\ref{Eq23a}), which characterizes the $S$-wave $D^0\bar
D^0\to\pi^+\pi^-$ annihilation at rest, are completely unknown. If
we naively put the inelastic scattering length $|\alpha''_{D^0\bar
D^0\to\pi^+\pi^-}|\approx1/(2m_{D^{*+}})\approx1/(4\,\mbox{GeV})$
(which is in dimensionless units $m_{\pi^+}|\alpha''_{D^0\bar
D^0\to\pi^+\pi^-}|\approx0.0347$), then $|g_{D^0\bar D^0\pi^+\pi^-}
/(8\pi)|^2$ is approximately equal to $\approx1.8$. We use this
value in further evaluations. It is clear that our rough estimate is
related to considerations about the $D^0\bar D^0$ annihilation
radius. An experiment will show whether this value is reasonable or
not. For comparison, we note that the tree $D^0\bar
D^0\to\pi^+\pi^-$ annihilation amplitude caused by the charged $D^*$
exchange leads to $|\alpha''_{D^0\bar D^0\to\pi^+\pi^-}|$, which is
about 15 times greater than our estimate, due to the large coupling
constant $g^2_{D^{*+}D^0\pi^+}/(4\pi)\approx5.6$ (see note
\cite{FN3}).

%--------------------------------------------------------------------------------
\begin{figure} %[!ht]
\begin{center}\includegraphics[width=8.2cm]{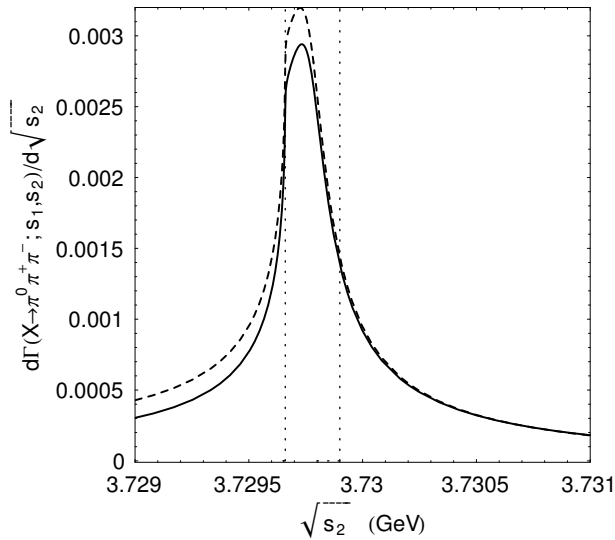}
\caption{\label{Fig-X-2} An example of the $\pi^+\pi^-$ mass
spectrum $d\Gamma (X\to\pi^0\pi^+\pi^-;s_1,s_2)/d\sqrt{s_2}$
constructed with the use of Eq. (\ref{Eq9a}) at
$\sqrt{s_1}=m_X=3.87169$ GeV and $g^2_A/(16\pi)=0.25$ GeV$^2$. The
solid curve corresponds to the sum of the diagrams in Figs. 2 and 3.
The dashed curve shows the contribution from the diagram in Fig. 2
only. The $\sqrt{s_2}$ values between which [according to Eq. (4)]
the amplitude of the $X(3872)\to(D^{*0}\bar D^0+\bar D^{*0}D^0)\to
\pi^0D^0\bar D^0\to\pi^0\pi^+\pi^-$ decay contains the logarithmic
singularity, in the hypothetical case of the stable $D^{*0}$ meson,
are shown by the dotted vertical lines. In so doing, the singularity
itself is located at $\sqrt{s_2}=3.72982$ GeV (see note \cite{FN4}).
}\end{center}\end{figure}
%--------------------------------------------------------------------------------

%--------------------------------------------------------------------------------
\begin{figure} %[!ht]
\begin{center}\includegraphics[width=8cm]{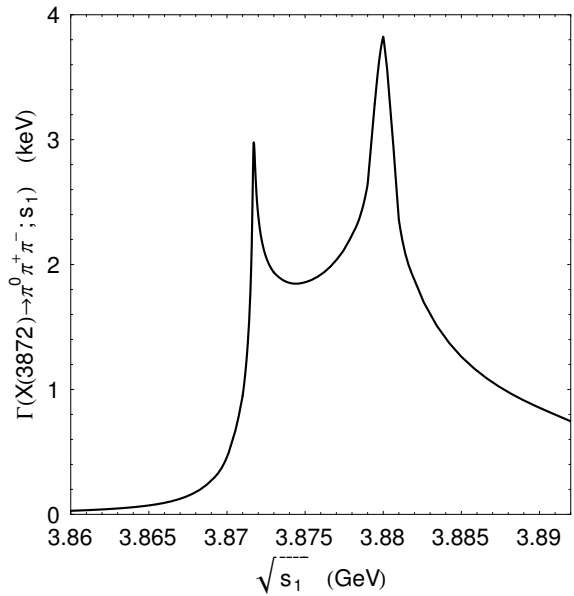}
\caption{\label{Fig-X-2} The width $\Gamma(X\to\pi^0\pi^+\pi^-;s_1)$
as a function of $\sqrt{s_1}$. The constructed example corresponds
to $g^2_A/(16\pi)=0.25$ GeV$^2$.}
\end{center}\end{figure}
%--------------------------------------------------------------------------------

Figure 4 shows an example of the $\pi^+\pi^-$ mass spectrum in the
decay $X(3872)\to\pi^0\pi^+\pi^-$, i.e., $d\Gamma(X\to\pi^0
\pi^+\pi^-; s_1,s_2)/d\sqrt{s_2}$ as a function of $\sqrt{s_2}$,
calculated with use of Eq. (\ref{Eq9a}) at $\sqrt{s_1}=m_X=3.87169$
GeV and the coupling constant of $X(3872)$ with the $D^{*0}\bar D^0$
channel $g^2_A/(16\pi)=0.25$ GeV$^2$ (other possible values for
$g^2_A/(16\pi)$ are discussed below). The integration
$d\Gamma(X\to\pi^0\pi^+ \pi^-; m^2_X,s_2)/d\sqrt{s_2}$ over
$\sqrt{s_2}$ in the region of 35 MeV wide, i.e., from
$m_X-m_{\pi^0}-0.035\mbox{ GeV}=3.70171\mbox{ GeV}$ to $m_X
-m_{\pi^0}=3.73671$ GeV, results in
$\Gamma(X\to\pi^0\pi^+\pi^-;m^2_X)\approx3$ keV. However, as can be
seen from Fig. 5, this is in fact the maximal value of the
$X(3872)\to\pi^0\pi^+\pi^-$ decay width in the $X(3872)$ resonance
region. The width $\Gamma(X\to\pi^0\pi^+\pi^-;s_1)$ is a sharply
changing function of $\sqrt{s_1}$. Two peaks in
$\Gamma(X\to\pi^0\pi^+\pi^-; s_1)$ located near the $D^{*0}\bar D^0$
and $D^{*+}D^-$ thresholds (see Fig. 5) are manifestations of the
logarithmic singularities in the amplitudes of the diagrams in Fig.
2 (the left peak) and in Fig. 3 (the right peak) \cite{FN5}. The
most important contribution to $BR(X\to\pi^0\pi^+\pi^-)$ [see Eq.
(\ref{Eq7a})] comes from the left peak. The right peak in $\Gamma
(X\to\pi^0 \pi^+\pi^-;s_1)$ practically does not work as it is
located far on the right tail of the $X(3872)$ resonance and its
contribution to $BR(X\to\pi^0 \pi^+\pi^-)$ is strongly suppressed by
the $X(3872)$ propagator module squared.

%--------------------------------------------------------------------------------
\begin{figure} %[!ht]
\begin{center}\includegraphics[width=8cm]{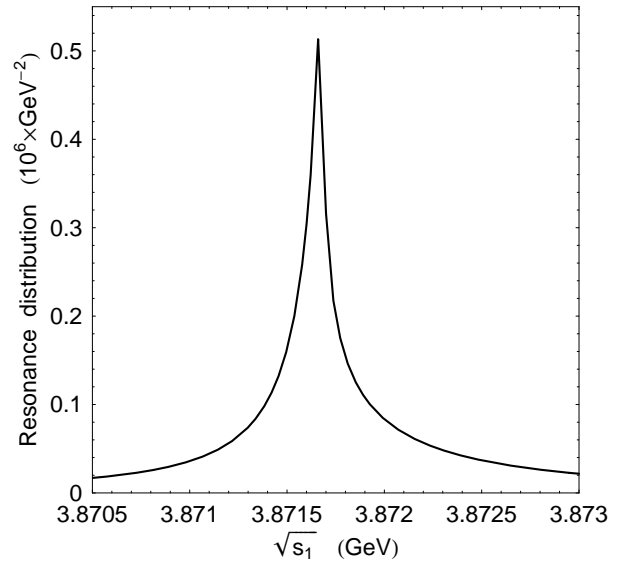}
\caption{\label{Fig-X-2} The resonance distribution
$2s_1/(\pi|D_X(s_1)|^2)$ at $g^2_A/(16\pi)=0.25$ GeV$^2$ and
$\Gamma_{non}=1$ MeV.}\end{center}\end{figure}
%--------------------------------------------------------------------------------

We now present numerical estimates for $BR(X\to\pi^0 \pi^+\pi^-)$
using as a guide the values of $g_A$ obtained in Refs.
\cite{AR14,AR16,A16}. Figure 6 shows an example of the resonance
distribution $2s_1/(\pi|D_X(s_1)|^2)$ calculated at $m_X=3.87169$
GeV \cite{PDG18}, $g^2_A/(16\pi)=0.25$ GeV$^2$, and $\Gamma_{non}=1$
MeV. Weighting with this distribution the energy dependent width
$\Gamma(X\to\pi^0\pi^+\pi^-;s_1)$ shown in Fig. 5, we find,
according to Eq. (\ref{Eq7a}), that for the above values of the
parameters $BR(X\to\pi^0\pi^+\pi^-)\approx5\times10^{-4}$. Estimates
for $BR(X\to\pi^0\pi^+ \pi^-)$ for different values of
$g^2_A/(16\pi)$ and $\Gamma_{non}$, which we vary in a fairly wide
but reasonable range, are given in Table I at $m_X=3.87169$ GeV
\cite{PDG18}.
%--------------------------------------------------------------------------------

\begin{table} [!ht]
\centering \caption{$BR(X((3872)\to\pi^0 \pi^+\pi^-)$ in units of
$10^{-4}$ for five values of $g^2_A/(16\pi)$ and three values of
$\Gamma_{non}$; $m_X=3.87169$ GeV.}\label{Tab1}\vspace*{0.1cm}
\begin{tabular}{|c|c|c|c|c|c|}
 \hline
 $g^2_A/(16\pi)$ (in GeV$^2$)  & = 0.1 & = 0.2 & = 0.25 & = 0.5 & = 1.0 \\
 \hline
 $\Gamma_{non}=0.5$ MeV       & 7.42 & 8.42 & 8.35 & 7.10 & 5.19 \\
 $\Gamma_{non}=1$ MeV         & 3.93 & 4.99 & 5.14 & 4.88 & 3.84 \\
 $\Gamma_{non}=2$ MeV         & 1.93 & 2.70 & 2.89 & 3.07 & 2.67 \\
 \hline
\end{tabular}
\end{table}
%--------------------------------------------------------------------------------

It is not yet clear whether the mass of the $X(3872)$ state lies
slightly above or slightly below the $D^{*0}\bar D^0$ threshold. The
$\pm0.17$ MeV uncertainty that the Particle Data Group \cite{PDG18}
indicates allows for both possibilities. Tables II and III show the
estimates for $BR(X\to\pi^0 \pi^+\pi^-)$ at the same values of
$g^2_A/(16\pi)$ and $\Gamma_{non}$ as in Table I  but for
$m_X=3.87169\pm0.00017$ GeV.

%--------------------------------------------------------------------------------

\begin{table} %[!ht]
\centering \caption{The same as Table I but for $m_X=3.87169
+0.00017$ GeV.}\label{Tab2}\vspace*{0.1cm}
\begin{tabular}{|c|c|c|c|c|c|}
 \hline
 $g^2_A/(16\pi)$ (in GeV$^2$)  & = 0.1 & = 0.2 & = 0.25 & = 0.5 & = 1.0 \\
 \hline
 $\Gamma_{non}=0.5$ MeV       & 6.45 & 6.97 & 6.82 & 5.63 & 3.94 \\
 $\Gamma_{non}=1$ MeV         & 3.76 & 4.60 & 4.68 & 4.30 & 3.27 \\
 $\Gamma_{non}=2$ MeV         & 1.93 & 2.64 & 2.80 & 2.89 & 2.45 \\
 \hline
\end{tabular}
\end{table}
%--------------------------------------------------------------------------------

\begin{table} %[!ht]
\centering \caption{The same as Table I but for $m_X=3.87169
-0.00017$ GeV.}\label{Tab3}\vspace*{0.1cm}
\begin{tabular}{|c|c|c|c|c|c|}
 \hline
 $g^2_A/(16\pi)$ (in GeV$^2$)  & = 0.1 & = 0.2 & = 0.25 & = 0.5 & = 1.0 \\
 \hline
 $\Gamma_{non}=0.5$ MeV       & 8.04 & 11.2 & 12.2 & 14.7 & 16.3 \\
 $\Gamma_{non}=1$ MeV         & 3.91 & 5.57 & 6.08 & 7.37 & 8.20 \\
 $\Gamma_{non}=2$ MeV         & 1.86 & 2.73 & 3.01 & 3.70 & 4.12 \\
 \hline
\end{tabular}
\end{table}
%--------------------------------------------------------------------------------

\section{Conclusion}

The above analysis shows that $BR(X(3872)\to\pi^0\pi^+\pi^-)$ can be
expected at the level of $10^{-3}$--$10^{-4}$. The dominant
contribution to $BR(X(3872)\to\pi^0\pi^+\pi^-)$ comes from the
production of the $\pi^+ \pi^-$ system in a narrow (no more than 20
MeV wide) interval of the invariant mass $m_{\pi^+\pi^-}$ near the
value of $2m_{D^0}\approx 3.73$ GeV. The $\pi^+\pi^-$ events with
such an invariant mass can serve as a signature of the decay
$X(3872)\to(D^{*0}\bar D^0+\bar D^{*0}D^0)\to\pi^0D^0\bar D^0\to
\pi^0\pi^+\pi^-$. \\[0.2cm]

The present work is partially supported by Grant No. II.15.1 of
fundamental scientific researches of the Siberian Branch of the
Russian Academy of Sciences, Grant No. 0314-2019-0021.

%--------------------------------------------------------------------------------
%\newpage
%--------------------------------------------------------------------------------

%--------------------------------------------------------------------------------


\begin{thebibliography}{99}
\bibitem{FN1}   M. Suzuki, Phys. Rev. D {\bf 72}, 114013 (2005).
Note that in the channel $X(3872)\to\omega J/\psi$ the $\omega$
meson is not observed literally, but the $\pi^+\pi^- \pi^0$ events
at the end of the left wing of the $\omega$ resonance which can else
contain the destructive or constructive interference with the
background.
\bibitem{FN3}  For the estimate we use the transverse $D^{*+}$ propagator
$\frac{-g_{\mu\nu}+\tilde{k}_\mu\tilde{k}_\nu/\tilde{k}^2}{
\tilde{k}^2-m^2_{D^{*+}}}$ that carries the unit spin off the mass
shell. So $\left|\frac{g_{D^0\bar D^0\pi^+\pi^-}}{8\pi}\right|^2=
\left|\frac{g^2_{D^{*+}D^0\pi^+}}{8\pi} \frac{s-u +(m^2_{D^0}-
m^2_{\pi^+})^2/t}{m^2_{D^{*+}}-t}\right|^2,$ where $s$, $t$, $u$ are
the usual Mandelstam variables for the reaction $D^0\bar D^0\to
\pi^-\pi^+$. At the $D^0\bar D^0$ threshold $s=4m^2_{D^0}$, $t=u=
-m^2_{D^0}+m^2_{\pi^+}$. Neglecting $m^2_{\pi^+}$, we get $\left|
\frac{g_{D^0\bar D^0\pi^+\pi^-}}{8\pi}\right|^2\approx \left|\frac{
g^2_{D^{*+}D^0 \pi^+}}{4\pi}\frac{2}{1+m^2_{D^{*+}}/ m^2_{D^0}}
\right|^2\approx27$. If the $D^{*+}$ propagator is $\frac{-g_{\mu
\nu}+\tilde{k}_\mu\tilde{k}_\nu/m^2_{D^{*+}}}{\tilde{k}^2-m^2_{
D^{*+}}}$, then $\left|\frac{g_{D^0\bar D^0\pi^+\pi^-}}{8\pi}
\right|^2\approx57$.
\bibitem{FN5} The contributions of the diagrams in Figs. 2 and 3 to
the width $\Gamma(X\to\pi^0\pi^+\pi^-; s_1)$ begin to cancel each
other at $s_1$ just below the $D^{*0}\bar D^0$ threshold and just
above the $D^{*+}D^-$ threshold. This practically does not occur
between the thresholds due to significantly different phases of the
neutral and charged intermediate states. The phase of the
$X\to\pi^0\pi^+\pi^-$ transition amplitude changes between the
$D^{*0}\bar D^0$ and $D^{*+}D^-$ thresholds by about 90$^\circ$,
just as is the case between the $K^+K^-$ and $K^0\bar K^0$
thresholds in the $a^0_0(980)-f_0(980)$ mixing amplitude \cite{AS17}
or in the amplitude of the decay $\eta(1405)\to f_0(980)\pi^0\to
\pi^+\pi^-\pi^0$ \cite{AS18}.
\bibitem{FN4}  The logarithmic singularities in the amplitude are at
$x^2_1+x^2_2+x^2_3+2x_1x_2x_3-1=0$, where $x_1=(s_1-m^2_{D^{*0}}
-m^2_{D^0})/(2m_{D^{*0}}m_{D^0})$, $x_2=(s_2-2m^2_{D^0})/(2m^2_{
D^0})$, $x_3=(m^2_{\pi^0}-m^2_{D^{*0}}-m^2_{D^0})/(2m^2_{D^{*0}}
m^2_{D^0})$ (see, for example, Refs. \cite{Lan59,FN64,Val64,Bay16,
AK1,AK2,AK3,AKS15,AS18}). Only one solution of this equation falls
into the region defined by Eqs. (\ref{Eq3a}) and (\ref{Eq4a}).
\end{thebibliography}
\end{document}